\documentclass[a4paper,12pt]{JHEP3}
\usepackage{longtable}
\usepackage{graphicx}
\usepackage{graphics}
\usepackage{keyval}
\usepackage{trig}
\usepackage{ulem}
\usepackage{dcolumn}
\usepackage{bm}
\usepackage{epsf,rotating,url}
\usepackage{subfigure,amssymb,amsfonts,amstext,amsgen,amsopn,amsxtra,indentfirst,lscape,times,rotating}
\def\beq{\begin{equation}}
\def\eeq{\end{equation}}
\def\bey{\begin{eqnarray}}
\def\eey{\end{eqnarray}}
\def\msun{M_\odot}

\def\kms{\, {\rm km \, s}^{-1} }

\def\grad{{\bf \nabla}}

\def\lcdm{{\Lambda}CDM}

\title{Modified Baryonic Dynamics: two-component cosmological simulations with light sterile neutrinos}
\author{G. W. Angus$^{1}$\thanks{E-mail: garry.angus@vub.ac.be}, A. Diaferio$^{2,3}$, B. Famaey$^{4}$, and K. J. van der Heyden$^{5}$ \\ 
$^{1}$Department of Physics and Astrophysics, Vrije Universiteit Brussel, Pleinlaan 2, 1050 Brussels, Belgium \\
$^{2}$Dipartimento di Fisica, Universit\`a di Torino, Via P. Giuria 1, I-10125, Torino, Italy \\
$^{3}$Istituto Nazionale di Fisica Nucleare, Via P. Giuria 1, I-10125, Torino, Italy\\
$^{4}$Observatoire Astronomique de Strasbourg, CNRS UMR 7550, France \\
$^{5}$Astrophysics, Cosmology \& Gravity Centre, Dept. of Astronomy, University of Cape Town, Private Bag X3, Rondebosch, 7701, South Africa \\
}
\abstract{ In this article we continue to test cosmological models centred on Modified Newtonian Dynamics (MOND) with light sterile neutrinos, which could in principle be a way to solve the fine-tuning problems of the standard model on galaxy scales while preserving successful predictions on larger scales. Due to previous failures of the simple MOND cosmological model, here we test a speculative model where the modified gravitational field is produced only by the baryons and the sterile neutrinos produce a purely Newtonian field (hence Modified Baryonic Dynamics). We use two component cosmological simulations to separate the baryonic N-body particles from the sterile neutrino ones. The premise is to attenuate the over-production of massive galaxy cluster halos which were prevalent in the original MOND plus light sterile neutrinos scenario. Theoretical issues with such a formulation notwithstanding, the Modified Baryonic Dynamics model fails to produce the correct amplitude for the galaxy cluster mass function for any reasonable value of the primordial power spectrum normalisation.}

\keywords{galaxy: formation – methods: N-body simulations – cosmology: theory
– dark matter – large scale structure of Universe}

\begin{document}

\date{\today}
\maketitle

\section{Introduction}
\protect\label{sec:intr}
Modified Newtonian Dynamics (MOND; \cite{milgrom83a}, and see \cite{famaey12} for a recent review) is a modification to gravity in the ultra weak-field regime. It has a remarkable ability to reproduce the dynamics of galaxies of all types, shapes and sizes (\cite{gentile11,ahd12,angus12,sanders13}) - with only a few notable caveats (\cite{gentile13,angus14}). On these scales, the standard cosmological model still struggles to reproduce the observed galaxy regularities and has a number of other problems (see e.g. \cite{walker14,kroupa10}. On large scales, like clusters of galaxies, the MOND approach does not work (\cite{aguirre01,sanders03,afb,afd10}). There is a missing mass problem for MOND in clusters of galaxies. There is also no clear way to describe cosmological phenomena like the anisotropies in the cosmic microwave background (CMB; \cite{spergel07,komatsu12,hinshaw12,sievers13}, but see \cite{skordis06}).

Approaching the missing mass problem from the other direction, cosmologists have found a satisfactory representation of the Universe at large scales using standard gravity (i.e. General Relativity). This comes through the combination of cold dark matter and dark energy ($\lcdm$), with baryons making up a mere 5\%. Invoking this combination allows the $\lcdm$ (or concordance) model to successfully match the acoustic peaks in the CMB (\cite{planck}). Furthermore, the observed distribution of matter on large scales (\cite{reiprich02,rines08,vikhlinin09b}) is well reproduced by N-body simulations of structure formation (\cite{press74,bardeen86,bond91,sheth02}). Something that would confirm the $\lcdm$ model would be cosmological N-body simulations that produce galaxies that replicate the properties of observed galaxies (\cite{walker14} and references therein). These observed properties of galaxies include the galaxy luminosity function (\cite{schechter76,cole01,bell03,blanton03}), the relative frequencies of the various 
Hubble types (\cite{kelvin14}), the colour bimodality (\cite{baldry04,balogh04,driver06}), and - perhaps most importantly - dynamical properties elegantly encapsulated by MOND. Two of the most significant of these dynamical properties are (i) the baryonic Tully-Fisher relation (\cite{mcgaugh00,mcgaugh05a}) and (ii) the scaling between the observed density profile of baryons and the inferred gravitational field (from a dynamical measure - \cite{mcgaugh05b}). This second property includes the requirement for dark matter (DM) to have a centrally cored distribution, rather than cusped (\cite{gentile04,governato10}). 

Although there are promising efforts to satisfy the aforementioned criteria (\cite{governato07,governato09,governato10,dutton12,stinson13a,stinson13b}), these last two requirements remain illusive. The most common belief is that more sophisticated modelling of the complex hydrodynamical feedback processes, from central black holes and massive stars, will allow the observations and predictions to be reconciled (\cite{sijacki12,vogelsberger13,kim14}).

Another approach is to assume MOND describes the dynamics of galaxies well and that we must blend a combination of MOND and DM together. The first attempt of this type came from \cite{sanders03} who added MOND and 2~eV active neutrinos together. This was not able to produce the measured acoustic peaks in the CMB, nor satisfy the requirement of DM in MOND clusters of galaxies (\cite{aszf,afb}) due to phase space constraints. \cite{angus09} suggested combining MOND with an 11~eV sterile neutrino, which was shown to be consistent with the CMB (under the ansatz that MOND is irrelevant at $z>10^3$) and the phase-space constraints from clusters of galaxies (\cite{afd10}). The added attraction of this idea (which was reviewed by \cite{diaferio14}) was that the 11~eV neutrinos would be fully thermalised, i.e. half of all quantum states would be occupied (just like the active neutrinos) - removing any fine tuning of their abundance. The other reason MOND plus sterile neutrinos is so attractive is that the free streaming 
properties of neutrinos on small scales means they would not influence galaxies. Thus, they would leave the impressive results of MOND in galaxies unblemished.

Using a purpose built MOND cosmological N-body code, \cite{ad11} showed this model was inconsistent with the observed cluster mass function - it produced too many high mass clusters and too few low mass clusters. \cite{angus13} further demonstrated this behaviour for MOND cosmological simulations to produce too many superclusters was not limited to 11~eV sterile neutrinos. In fact, all masses of sterile neutrinos were ruled out, regardless of (i) the value of the MOND acceleration constant, (ii) the redshift at which MOND is ``switched on'', (iii) the normalisation of the initial conditions and (iv) the interpolating function. Other factors like the equation of state of dark energy were also shown to be ineffective.

In summary it does not appear to be possible to form the correct halo mass function in standard MOND from any sterile neutrino initial conditions that grew from an initially Harrison-Zel'dovich power spectrum under GR until $z\sim200$. So if MOND is the correct description of gravitational dynamics on galaxy scales, then either the initial conditions are not as described above and yet conspire to produce the correct CMB angular power spectrum, or perhaps MOND does not affect the sterile neutrinos. This is important because although galaxies require MOND to form (and stably exist) without CDM, the clusters clearly do not require MOND at all and one should not ignore how well Newtonian gravity reproduces the cluster mass function.

At minimum, the $\lcdm$ model gives the correct cluster scale halo mass function at $z=0$, whether some additional boost to gravity is required to form the clusters early enough has been discussed in the literature (\cite{mullis05,bremer06,jee09,jee11,rosati09,brodwin10,brodwin12,foley11}). MOND has a double negative effect on the cluster mass function if it influences the sterile neutrinos. Not only does it facilitate more rapid growth and the formation of much larger and denser structures than in Newtonian gravity, but these more massive halos now have MOND gravity meaning their dynamical masses are further enhanced, causing poorer agreement with the data. This result might suggest that if the MOND gravitational field is not produced by the sterile neutrinos (meaning only a Newtonian gravitational field is produced by them), but is only produced by the baryons, that it will have a positive influence on the halo mass function. Below we address how significant this influence is.

There is another factor to consider here which rules out the idea of MOND not being activated until some low redshift. That is that galaxies in MOND must form without the aid of a dark matter halo (cold, warm or hot) and galaxy formation without dark matter (if it is possible at all) is only possible with the added benefit of stronger than Newtonian gravitational attraction between the baryons. Thus, if MOND was not in effect until $z=1$, then galaxies would not {\it begin} to form until then and galaxies are clearly formed long before this.

The simple options to blend MOND with DM are therefore ruled out. There are also more involved ideas, such as the tantalising theory of dipolar dark matter \cite{blanchet08,blanchet09} and bimetric MOND (\cite{milgrom10a,milgrom13}), which require further investigation. However, a simpler framework may still exist with sterile neutrinos and MOND.

In the traditional framework of MOND, there exists no dark matter in galaxies. Therefore, it is assumed that only the baryons produce a modified gravitational field. In the extended frameworks, where MOND is blended with some species of neutrinos, it has always been assumed that the baryons and the neutrinos contribute to the modified gravitational field. On the other hand, there do exist modified theories of gravity (completely different to MOND) where the dark matter has a mutual fifth-force interaction, to which the baryons do not participate (e.g. \cite{nusser05,farrar07}). Here we go the other way, and suggest the baryons produce a modified gravitational field, but the sterile neutrinos do not. This can in principle retain all the benefits of adding sterile neutrinos to MOND - such as addressing the acoustic peaks of the CMB and solving the missing mass problem in MOND clusters. It also might evade the problems engendered when the sterile neutrinos produce a modified gravitational field - like 
overproducing massive clusters. Furthermore, in the original MOND + sterile neutrino framework, the effectiveness of MOND had to be ``switched off'' (i.e. the acceleration constant of MOND had to be substantially decreased) prior to recombination in an ad hoc way. This was to avoid MOND altering the shape of the CMB acoustic power spectrum. In a model where the sterile neutrinos do not produce a modified gravitational field, this may not be necessary.

In this article we run MOND cosmological simulations in a framework where the modified gravitational field is produced only by the baryons, not the sterile neutrinos. We present the framework and describe the simulation setup in \S2, show the results of our simulations in \S3 and in \S4 we give our conclusions.

\section{Background and Method}
\protect\label{sec:code}

The Quasi-Linear formulation of MOND (QUMOND; \cite{milgrom10}) requires solution of a modified version of the Poisson equation. Specifically, the ordinary Poisson equation for cosmological simulations

\beq
\protect\label{eqn:qumond1}
\nabla^2\Phi_N=4\pi G (\rho - \bar{\rho})/a,
\eeq
is solved to give the Newtonian potential, $\Phi_N$, at scale factor $a$, from the ordinary matter density $\rho$ that includes baryons and neutrinos. This would also include cold dark matter if there was any in our model. The QUMOND potential, $\Phi$, is found from the Newtonian potential as follows
\beq
\protect\label{eqn:qumond2}
\nabla^2\Phi=\grad \cdot \left[ \nu(y) \grad\Phi_N \right],
\eeq
and $y=\nabla\Phi_N/a_o a$, with $a_o$ being the MOND acceleration constant chosen here to be $3.6~(\kms)^2pc^{-1}$. The interpolating ($\nu$) function is parametrised as per \cite{famaey12} Eqs. 51 and 53 where
\beq
\protect\label{eqn:nualp}
\nu_{\alpha}(y)=\left[{1+(1+4y^{-\alpha})^{1/2}\over 2}\right]^{1/\alpha},
\eeq
and $\alpha=1$ is the so-called simple $\nu$-function and $\alpha=2$ is the standard $\nu$-function.

The specifics of how to solve Eqs~\ref{eqn:qumond1} \& \ref{eqn:qumond2} are also explained in AD11, but we review the main points here.

The code we wrote is particle-mesh based, with a grid-mesh that has a limit of $257$ cells in each dimension. In all our simulations we use $256$ particles per dimension. Particle-mesh solvers are required to handle the non-linear MOND Poisson equation. Direct or tree-code methods fail because in MOND we cannot co-add particle gravities. The full MOND Poisson equation must be solved because the trivial MOND equation of spherical symmetry does not satisfy the conservation laws.

Our one-component simulations are described by the following procedure

\begin{itemize}
\item The particle positions and velocities are read in and the density of the particles is assigned to the various cells with the cubic cloud-in-cell method.
\item We use multi-grid methods (see Numerical Recipes \S19.6) to solve the Poisson equation to find the Newtonian potential (Eq~\ref{eqn:qumond1}). A 3D black-red sweep to update the cells with the new approximation of the potential in that cell and we iterate until we have fractional accuracy of $10^{-10}$.
\item We take the divergence of the vector in the square brackets of Eq~\ref{eqn:qumond2} which gives us the source of the QUMOND potential.
\item We then repeat the Poisson solving step with the new source density to give the QUMOND potential, $\Phi$, which we take the gradient of to find the gravity at each cell.
\item We then interpolate to each particle's position to find the appropriate gravity and move each particle with a second order leapfrog.
\item The procedure repeats from the second stage until the simulation reaches $z=0$.
\end{itemize}

\subsection{Modified Baryonic Dynamics}
\protect\label{sec:mm}
The modification of gravity we propose is the following: the Newtonian potentials of the sterile neutrinos and baryons are found by solving the Newtonian Poisson equation

\bey
\protect\label{eqn:qumond1a}
\nonumber \nabla^2\Phi_{N,\nu_s}&=&4\pi G (\rho_{\nu_s} - \bar{\rho_{\nu_s}}))/a\\
\nabla^2\Phi_{N,b}&=&4\pi G (\rho_{b} - \bar{\rho_{b}})/a.
\eey
The gravitational field produced by the sterile neutrinos is not modified and is thus $-\grad\Phi_{N,\nu_s}$. On the other hand, the gravitational field produced by the baryons is found from the analogue of the QUMOND equation (Eq~\ref{eqn:qumond2}), which is

\beq
\protect\label{eqn:qumond2a}
\nabla^2\Phi_b=\grad \cdot \left[ \nu(y) \grad\Phi_{N,b} \right],
\eeq
where crucially 
\beq
\protect\label{eqn:nuf2}
y={|\grad\Phi_{N,b}+\grad\Phi_{N,\nu_s}| \over a_o a}.
\eeq

The total modulus of the gravitational field experienced by any particle, sterile neutrino or baryon, is

\beq
\protect\label{eqn:qumond3a}
\grad\Phi=\grad\Phi_{N,\nu_s}+  \nabla\Phi_{b}.
\eeq
Given that a modified gravitational field (relative to Newtonian gravity) is only produced by the baryons, we refer to this model as Modified Baryonic Dynamics (MBD). Note that, at this point, it is a phenomenological approach at the classical level: whether a Lagrangian producing these equations of motion can be found (both at the classical and covariant level) could be the subject of further work, if necessary. It is plausible that such a Lagrangian will require couplings of the baryons and sterile neutrinos to two different metrics, thereby leading to a violation of the weak-equivalence principle, which would have additional interesting consequences in their own right (see, e.g., \cite{alimi08}).

\subsection{Initial Conditions and Simulation Setup}
\protect\label{sec:ics}
In \cite{ad11} and \cite{angus13} we made use of the original \verb'COSMICS/GRAFICS' package of \cite{bert95} to generate our initial conditions. We chose to input our own transfer functions using the massive neutrino parametrisation of \cite{abazajian06} (their equations 10-12) and the resulting linear matter power spectra were plotted for a sample of neutrino masses in \cite{angus13} Fig~1.

We exploit the new CMB results provided by the PLANCK mission (\cite{planck}) which identifies $\Omega_b$, $\Omega_{\nu_s}$, $\Omega_{\Lambda}$, $h$, $n_s$)=(0.049, 0.267, 0.683, 0.671, 0.962).

We use the quadrupole temperature, $Q_{rms-PS}$, as a free parameter to fit the amplitude of the cluster mass function in MBD.

The CMB quadrupole, $Q_{rms-PS}$, is used to normalise the initial power spectrum of perturbations in the same way as $\sigma_8$ typically is for CDM simulations, because one cannot use linear theory in MOND to estimate $\sigma_8$ at $z = 0$.

In order to exploit the MBD formulation we use the \verb'GRAFICS-2' package of \cite{bert01} which generates two sets of particles: baryons and sterile neutrinos (or any other DM particle one chooses), giving $256^3$ for each species. The masses of the sterile neutrino and baryonic particles are found from $m=1.4\times10^{11}\Omega_{i}(L_{box}/N_p)^3\msun$, where $\Omega_i$ should be replaced with either $\Omega_{\nu_s}$ or $\Omega_b$. It is then perfectly straight-forward to code the above expressions.

We then proceed as follows:

\begin{itemize}
 \item We read in the positions of the sterile neutrinos, compute their density ($\rho_{\nu_s}$) on the grid and then solve for their Newtonian potential ($\Phi_{N,\nu_s}$, as per Eq~\ref{eqn:qumond1a}) using finite differencing and multi-grid methods as described in AD11.
 \item We read in the positions of the baryons, compute their density on the grid ($\rho_{b}$) and then their Newtonian potential ($\Phi_{N,b}$, as per Eq~\ref{eqn:qumond1a}).
 \item We derive $\grad \cdot \left[ \nu(y) \grad\Phi_{N,b} \right]$, the QUMOND source density for the baryons, which uses an argument for the interpolating function as per Eq~\ref{eqn:nuf2}.
 \item We find the QUMOND potential for the baryons $\Phi_b$, using Eq~\ref{eqn:qumond2a}.
 \item We add the QUMOND potential for the baryons together with the Newtonian potential for the sterile neutrinos (as in Eq~\ref{eqn:qumond3a}) to give the total potential, from which the gravitational field is derived to move both sets of particles.
\item We then interpolate to each particle's position to find the appropriate gravity and move each particle with a second order leapfrog.
\item The procedure repeats from the second stage until the simulation reaches $z=0$.

\end{itemize}

\section{Results}
Using two component simulations with sterile neutrinos, the baryons almost exactly trace the sterile neutrino distribution, which is not surprising. As with the original single component MOND simulations (\cite{angus13}), when computing the mass function, we use the Newtonian equivalent mass of halos - that is the dynamical mass our MOND halos would appear to have if probed using Newtonian dynamics. With MBD it is merely a case of separately running the AMIGA halo finder (\cite{gill04,knollmann09}) on both the baryons and sterile neutrinos (which gives identical halo numbers and halo particle mass distributions for our cluster sized halos) and calculating the Newtonian equivalent mass, $M_m(r)=M_b(r)\nu\left( {GM_{\nu+b}(r) \over r^2a_o}\right)+M_{\nu}(r)$ since the modification to gravity only affects the baryonic gravitational field. $M_b$ and $M_{\nu}$ represent the particle mass profiles of baryons and sterile neutrinos respectively. From here it is merely a case of finding the 
radius where $M_m(r)$ encloses an average density of matter 200 times the critical density and building the mass function. We only consider the mass bins of our mass function to be relevant if more than 8 halos are found within the bins of 0.233 dex per decade of mass.

We ran 12 two component simulations with different initial normalisations through the CMB quadrupole and different sterile neutrino masses. For each set of parameters, we ran two simulations: one with a 128~Mpc/h box and the other with a 256~Mpc/h box. We did this because we do not have adaptive mesh refinement built into our code. Therefore, in the simulations with the larger box, we have insufficient spatial resolution which leads to a suppression of the mass function. On the other hand, with the smaller boxes we naturally form fewer halos and have poor statistics. This issue is described in more detail in \cite{angus13}. Since we have twice the spatial resolution in the 128~Mpc/h simulations, we use the amplitude of their mass functions to re-normalise the 256~Mpc/h simulations by matching their amplitude to the 128~Mpc/h boxes at a single mass bin. From comparison with theoretical mass functions (\cite{angus13}), we know the 128~Mpc/h box simulations produce roughly the correct mass function normalisation.

\begin{figure*}
\includegraphics[angle=0,width=15.0cm]{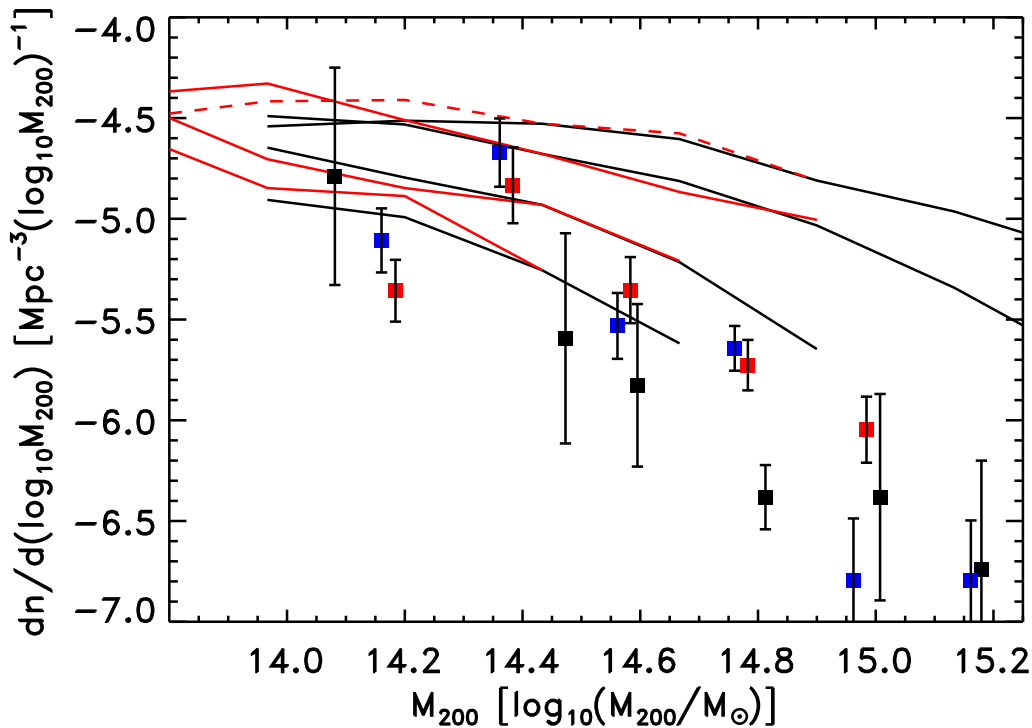}
\caption{Here we plot the mass functions from simulations with the same sterile neutrino mass of 100~eV, but with different CMB quadrupole normalisations (from the bottom curve upwards: $Q_{rms-PS}$=4, 5, 9 and 17~$\mu K$. The red lines represent the 128~Mpc/h simulation boxes and the black lines represent the re-normalised mass functions of the 256~Mpc/h boxes. In this figure, the 256~Mpc/h box mass functions are increased to have the same amplitude as the 128~Mpc/h box simulations at $M_{200}=10^{14.4}\msun$.}
\label{fig:norm}
\end{figure*}

In Fig~\ref{fig:norm} we plot the mass functions from simulations with the same neutrino mass of 100~eV, but various normalisations through the CMB quadrupole. The red lines represent the 128~Mpc/h boxes and the black lines represent the renormalised mass functions of the 256~Mpc/h boxes. We only plot the curves for mass bins that have eight or more halos (as mentioned previously). The four normalisations used, in increasing amplitude, represent $Q_{rms-PS}$= 4, 5, 9 and 17$~\mu K$. Clearly $Q_{rms-PS}<5~\mu K$ is required to have a mass function amplitude which is comparable to the data (for a 100~eV sterile neutrino).

In general, the mass functions found using different normalisations, through the CMB quadrupole ($Q_{rms-PS}$), in pure MOND showed no variation (Fig 7, \cite{angus13}) i.e. the mass functions had the same amplitude at $z=0$ regardless of initial normalisation. The mass functions found with MBD clearly vary with normalisation. 

\begin{figure*}
\includegraphics[angle=0,width=15.0cm]{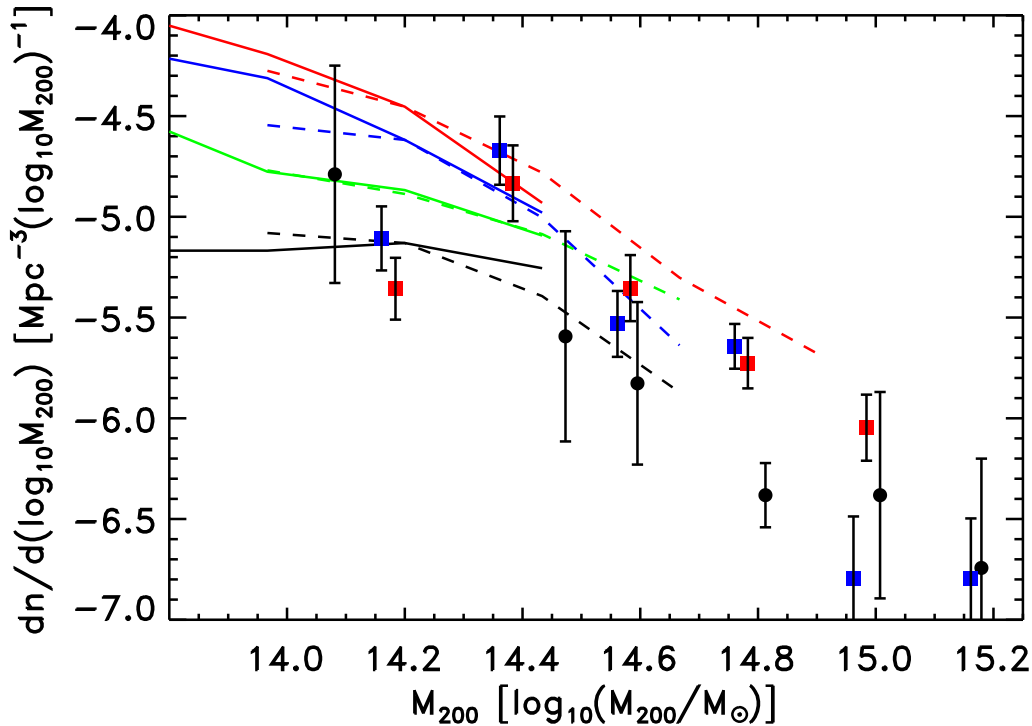}
\caption{Here we plot the mass functions from simulations with the same CMB quadrupole normalisation, but different sterile neutrino masses (starting from the bottom curve): 50, 100, 150 and 300~eV. The red lines represent the 128~Mpc/h simulation boxes and the black lines represent the re-normalised mass functions of the 256~Mpc/h boxes. In this figure, the 256~Mpc/h box mass functions are increased to have the same amplitude as the 128~Mpc/h box simulations at $M_{200}=10^{14.2}\msun$.}
\label{fig:mass}
\end{figure*}

In Fig~\ref{fig:mass} we plot the mass functions from simulations with the same normalisation through the CMB quadrupole, of $Q_{rms-PS}=4.5~\mu K$, but various neutrino masses 50, 100, 150 and 300~eV. We find that the lower the sterile neutrino mass, the lower the amplitude of the mass function and better the agreement with the observed cluster mass function. We only tested sterile neutrino masses of 50, 100, 150 and 300~eV, but using a mass of 30~eV or lower would not allow a fit to the cluster mass function. However, regardless of mass, a very low normalisation is always required.

To test if there was any sensitivity to the $\nu$ function (Eq~\ref{eqn:nualp}) employed, we ran two further MBD simulations using $\alpha=2$ (the standard $\nu$ function), where the other MBD simulations used $\alpha=1$ (the simple $\nu$ function). We chose $Q_{rms-PS}$=4.5 and 12.5~$\mu K$ and the result was that there is very little difference between simulations with $\alpha=1$ and $\alpha=2$. This means using any reasonable $\nu$ function in MBD will still require a low normalisation for the initial spectrum of perturbations, through the CMB quadrupole.

\subsection{The significance of the CMB quadrupole}

The measured CMB quadrupole value is related to the multipole moment $l=2$ coefficient, $C_2^{TT}$, as $Q_{rms-PS}=\left(5C_2^{TT}\over4\pi\right)^{1/2}$. \cite{bennett11} and \cite{bennett13} plot likelihood versus quadrupole value in their Figs~3 and 38 respectively. The allowed range at one sigma has shifted towards lower values in the latest paper and currently the lower one sigma confidence level is $C_2^{TT}\approx 50\mu K^2$ (the maximum likelihood is around $160\mu K^2$). This leads to $Q_{rms-PS}\approx4.5\mu K$ at one sigma and roughly $8~\mu K$ at maximum likelihood.

The fitted CMB quadrupole we require to be consistent with the galaxy cluster mass function is $Q_{rms-PS}<5\mu K$, which is roughly one sigma from the observed quadrupole. 

The measured quadrupole is obviously separate from the fitted quadrupole. A quadrupole value is fitted to the CMB angular power spectrum.

In the $\lcdm$ model, a fitted quadrupole of around $21.4~\mu K$ (\cite{bennett13}) is required so that the amplitude of the theoretical CMB power spectrum at mulitpoles $l>2$ matches the observed spectrum. There should also be consistency between the theoretical value of the quadrupole required to fit the CMB and other cosmological probes such as baryonic acoustic oscillations, galaxy clustering and the galaxy cluster mass function i.e. $\sigma_8$ should be the same for each probe. The theoretical quadrupole is between one and two sigma larger than the measured one. However, consistency with the observed one can be achieved by models that suppress the theoretical one (due to inflationary theories, alternative gravity theories, etc) or increase the observed quadrupole (due to foreground modelling).

In the MBD case, we also expect the fitted quadrupole used to normalise the CMB ($21.4~\mu K$ as per the $\lcdm$ value) and found from matching the galaxy cluster mass function to be the same. Clearly, the galaxy cluster mass function requires $Q_{rms-PS}<5\mu K$ and the CMB requires $Q_{rms-PS}\approx21.4\mu K$, so they are hugely discrepant. This is far more serious than a single, consistent fitted quadrupole overestimating the measured one - as in the $\lcdm$ case.

Therefore, our basic assumption that the CMB power spectrum would be the same in MBD as in $\lcdm$ must be revised. Instead, to check the validity of the MBD model, a full treatment of CMB anisotropies is required and if a model is found which matches the CMB power spectrum, its normalising CMB quadrupole amplitude should be consistent with the one used to fit the galaxy cluster mass function.

\section{Conclusion}
In previous works (\cite{ad11,angus13}) we tested the hypothesis that combining MOND with sterile neutrinos could produce the observed mass function of clusters of galaxies. We found it could not. This meant that, if MOND is the correct description of weak-field gravitational dynamics on galaxy scales, then either the whole cosmology and/or the initial conditions are not as described above (see, e.g., Sect. 9.2 of \cite{famaey12} for a discussion in the context of covariant MOND theories), and yet would conspire to produce the correct CMB angular power spectrum, or perhaps MOND does not affect the sterile neutrinos in the same way as the baryons.

For this reason, we propose that the modified gravitational field of MOND is only produced by the baryons. In this scenario, the sterile neutrinos produce a Newtonian gravitational field. However, the baryons - subject to the total Newtonian gravitational field (both from the sterile neutrinos and baryons) - produce a MOND-like gravitational field. 

There are similarities between our MBD model and many of the fifth force models referenced in \cite{peebles10}. For instance \cite{farrar07} employed a long range scalar field, generated only by the dark matter (not by the baryons), to increase the mutual acceleration of the two clusters comprising the colliding bullet cluster. In our MBD model, only the baryons generate a stronger than Newtonian gravitational attraction, whilst the sterile neutrinos produce a purely Newtonian gravitational field.

Our MBD model is far more conducive to producing a halo mass function that satisfies observational constraints than our previous attempts. Nevertheless, it requires fine tuning the normalisation of the primordial power spectrum, as is often done, through the CMB quadrupole, $Q_{rms-PS}$.

It was found that a normalisation of $Q_{rms-PS} < 5~\mu K$ was required to have an adequate match and a lower sterile neutrino mass was preferred over a higher one.     

The CMB quadrupole required is around 1~$\sigma$ lower than the measured value from CMB observations (\cite{bennett13}). More significantly, this is much lower than the normalisation required to fit the CMB in $\lcdm$ ($21.4~\mu K$). This is crucial because we previously adopted the $\lcdm$ fits to the CMB as a basis for any MOND plus sterile neutrino theory, under the assumption the two models would give an identical CMB power spectrum.

We conclude that the simple model considered here is not viable, and that more involved MOND models should be considered if the MOND phenomenology is indeed derived from a fundamental modification of the Lagrangian of Nature rather than an emergent phenomenon at galaxy scales. These theories must include at least one new degree of freedom corresponding to a non-trivial dark matter fluid with a non-trivial coupling to baryons. Interesting possibilities in this vein include dipolar dark matter  \cite{blanchet08,blanchet09,blanchet14}  or BIMOND with twin matter \cite{milgrom10a,milgrom13}, as well as other possible frameworks still to be conceived.

\section{Acknowledgements} GWA's research is supported by the FWO - Vlaanderen.

\bibliography{mbd}

\end{document}